\documentclass[%
 reprint,
 amsmath,amssymb,
 aps,
]{revtex4-2}
\usepackage{physics}
\usepackage{bbold}
\usepackage{amsmath}
\usepackage{latexsym}
\usepackage{graphicx}
\usepackage{amsthm}
\usepackage{hyperref}
\usepackage{lipsum}
\usepackage[english]{babel}
\usepackage{slashed}
\usepackage[compat=1.1.0]{tikz-feynman}
\usepackage{appendix}
\usepackage{subfig}
\usepackage{multirow}

\theoremstyle{plain}

\begin{document}

\preprint{APS/123-QED}

\title{Tree-level entanglement in Quantum Electrodynamics} % at Tree Level / in Quantum Field Theory / in Particle Scatterings / in Quantum Electrodynamics

\author{Samuel Fedida}
\email{samuel.fedida.18@ucl.ac.uk}
\author{Alessio Serafini}
\email{serale@theory.phys.ucl.ac.uk}
\affiliation{Department of Physics \& Astronomy, University College London, \\ Gower Street, London WC1E 6BT, UK}

\date{\today}

\begin{abstract}
    We report on a systematic study on the entanglement between helicity degrees of freedom generated at tree-level in quantum electrodynamics two-particle scattering processes. We determine the necessary and sufficient dynamical conditions for outgoing particles to be entangled with one another, and expose the hitherto unknown generation of maximal or nearly maximal entanglement through Bhabha and Compton scattering. Our work is an early step in revisiting quantum field theory and high-energy physics in the light of quantum information theory.
\end{abstract}

\maketitle

\section{Introduction}

%Theory (PERES-HORODECKI, time-evolution of rho from S-matrix, ...)

Quantum Information Theory (QIT), i.e. the study of information processing based on Quantum Mechanics (QM) rather than classical laws, has brought to the fore a number of interesting facets whereby QM deviates dramatically from its classical predecessor. One such information-related non-classical aspect is certainly quantum entanglement, that is, the occurrence of correlations in quantum systems that cannot be reproduced in classical systems, leading to the violation of Bell inequalities.
As was often remarked in various forms, quantum entanglement is {\em the} fundamental distinguishing feature that sets quantum theory apart from classical phenomena, and has drawn an enormous amount of attention over the past 25 years, both in blue sky research and as a key ingredient in quantum technologies, central to quantum teleportation \cite{Ren2017}, cryptography \cite{Yin2020}, computation \cite{Jozsa2003} and metrology \cite{Demkowicz2014}.
On the other hand, the most fundamental expression of quantum mechanical dynamical laws is given by Quantum Field Theory (QFT), the relativistic formulation of quantum dynamics, where particles and interactions are depicted via fields spanning through all of spacetime. 

%In particular, one QFT of interest is Quantum ElectroDynamics (QED), which models the electromagnetic interaction on a quantum level.

Notwithstanding the centrality of the notion of entanglement, the role of entanglement in QFT is still relatively little studied. Typically, past research has focused on scaling laws,
for bosonic and more general field theories \cite{Eisert2010} as well as conformal field theories \cite{Calabrese2004} and black holes, bearing significant foundational consequences in the context of the ADS-CFT correspondence \cite{Ryu2006,Jensen2013,Maldacena2013} and quantum gravity \cite{Cao2018}. 

In the more specific context of high-energy physics, entanglement has mostly been considered through the study of Bell inequalities as tests for discrete fundamental symmetries, notably time-reversal, charge conjugation and parity. Kaonic systems are usually examined for such theoretical explorations \cite{Benatti1998,BertlmannJan2001,BertlmannJuly2001,Bertlmann2006,Lello2013}, and are used as probes of special relativity via the experimental tests of the CPT theorem at B- and $\Phi$-factories \cite{Amelino2010,Bernabeu2011}.
Spin correlations and the violation of Bell-like inequalities in Bhabha scattering have been discussed in \cite{yongram08,beck22}.
Entanglement has also been investigated in Quantum ChromoDynamics (QCD) \cite{Afik2022}, for gluon pairs \cite{Seki2014, Beane2019} and deep inelastic scattering \cite{Kovchegov2012, Kharzeev2017}, and in neutrino oscillations \cite{Go2004,Blasone2008,Blasone2009,Blasone2010,Blasone2014a,Blasone2014b,Banerjee2015}. 
The von Neumann entropy of entanglement has been evaluated in some scattering processes as well, in $\phi^4$ theory \cite{Seki2015}, Quantum ElectroDynamics (QED) \cite{Fan2018}, and in QFT more generally \cite{Peschanski2016}. 
A related analysis of entanglement in relativistic quantum mechanics has been performed in \cite{Pachos2003}, and it was found that the non-local nature of entanglement is consistent with relativistic causality, which has been illustrated in the context of QED.

A more systematic inquiry into the generation of entanglement in QED processes at tree-level has been carried out more recently by Cervera-Lierta and coworkers \cite{Cervera-Lierta2017,Cervera-Lierta2019}, with an emphasis on the conditions to generate maximal entanglement between helicity degrees of freedom, partly extended to the electroweak sector too, which is then speculated to form a fundamental `it from bit' principle for fundamental interactions.

In the present paper, we will expand on Cervera-Lierta {\em et al.}'s work by providing a detailed, comprehensive analysis of the entanglement generated at all energies and for pure and mixed final states in QED two-particle scatterings at tree-level, for arbitrary initial mixtures of helicity states. Thus, we will shed light on hitherto unknown generation mechanisms involving only dominating $t$- or $u$-channels.
Further to probing the non-trivial structure of quantum correlations in ubiquitous, elementary electromagnetic interactions in their most primitive description, an endeavour of fundamental significance, our results may also indirectly contribute to the ongoing debate on the certification of the quantum nature of fundamental forces, such as gravity.

%%%%%%%%%%%%%%%%%%%%%%%%%%%%%%%%%%%
\section{Setting and preliminaries}

We will consider the scattering of two incoming particles into two outgoing particles interacting through QED vertices, followed by an arbitrarily sharp filtering of the outgoing particles in momentum space, without resolving their internal (helicity or polarisation) degrees of freedom. The result of such an idealised measurement is a four-dimensional final density matrix (i.e., a two-qubit state, each qubit being embodied by the helicity of one of the outgoing particles), which will be reconstructed to the lowest perturbative order. Hence, the necessary and sufficient Peres-Horodecki separability criterion, and the related evaluation of the logarithmic negativity, will be applied on the final state to qualify and quantify its entanglement.

Note that, although the momentum filtering we shall assume is idealised, the entanglement we will calculate between the helicity degrees of freedom at given momenta is still a fundamental property of the output state of the quantum field upon scattering. In practice, such filtering would correspond to collimating the output particles' trajectories.

%%%%%%%%%%%%%%%%%%%%%%%%%%%%%%%%%%%%%
\subsection{Derivation of the density matrix from the scattering matrix elements}

In order to study quantum entanglement we will need to access all the off-diagonal elements of the final helicity density matrix (its `coherence', so to speak) 
after scattering and momentum filtering. This is hardly ever considered 
in traditional approaches to quantum field theory, whose ultimate 
focus is on scattering amplitudes, but can of course be accomplished 
in terms of Feynman diagrams \cite{Stanton1971,Hagston1980}.

The time-evolution of a density matrix is given by \begin{equation}
    \rho_t = U(t,t_0) \rho_{t_0} U^\dagger(t,t_0)
\end{equation} where $U(t,t_0)$ is the time-evolution operator. When we study scatterings in QFT, the initial system is considered at time $t = -\infty$ whilst the system after the scattering is at $t = +\infty$, with the S-matrix operator defined as $
    S = \lim_{t \rightarrow +\infty} \lim_{t_0 \rightarrow -\infty} U(t,t_0)
$, so through a scattering, a density matrix evolves as  \begin{equation} \label{evol}
    \rho_{+\infty} = S \rho_{-\infty} S^\dagger = 
    \sum_{\lambda}p_{\lambda} S|p,\lambda\rangle \langle p,\lambda|S^{\dag} \, ,
\end{equation} 
where $p$ stands for a configuration of momenta while $\lambda$ stands for 
a configuration of helicity eigenstates (for both particles) and we have 
assumed a generic diagonal mixture 
$\sum_{\lambda}p_{\lambda} |p,\lambda\rangle \langle p,\lambda|$
as the initial state, with $\sum_{\lambda}p_\lambda=1$.

Momentum filtering along momenta $q$ is described by applying 
the POVM element $\Pi_q=\sum_{\eta}|q,\eta\rangle\langle q,\eta|$ on such a final state, obtaining the output, two-qubit helicity state 
\begin{align}
    \rho_{out} &= \frac{\Pi_q \rho_{\infty} \Pi_q}{{\rm Tr}[\Pi_q \rho_{\infty} \Pi_q]}\nonumber\\&=\frac{ \sum_{\lambda,\eta,\eta'}p_{\lambda} 
    |q,\eta\rangle\langle q,\eta|S|p,\lambda\rangle \langle p,\lambda|
    S^{\dag}|q,\eta'\rangle\langle q,\eta'|}{ \sum_{\lambda,\eta}p_{\lambda} 
    \langle q,\eta|S|p,\lambda\rangle \langle p,\lambda|
    S^{\dag}|q,\eta\rangle} \label{rhozzo}\\
    &= \sum_{\lambda,\eta,\eta'} \rho_{\eta,\eta'}
    |q,\eta\rangle\langle q,\eta'| \, . \nonumber 
\end{align}
Up to normalisation, which can always be restored \textit{a posteriori},
this expression yields the final filtered state in terms of S-matrix elements 
$S_{qp}=\langle q,\eta|S|p,\lambda\rangle$, which correspond to Feynman diagrams and can be evaluated through Feynman rules at a certain order. 
In keep with the standard notation, we shall in the following  
report scattering amplitudes $\mathcal{M}_{fi}$, related to $S_{qp}$ via
    $S_{qp} = i (2\pi)^4 \delta^4(p-q) \mathcal{M}_{fi}$. Note that 
    by working in the centre of mass frame only two dynamical parameters (a scattering angle and the magnitude of incoming three-momentum) will determine the scattering amplitudes and the ensuing entanglement.

In what follows, we shall adopt a standard ``left/right'', $L, R$, notation for the helicity eigenstates, whereby the helicity labels above will take the four values $LL$, $LR$, $RL$ and $RR$. We shall also adopt the notations $|LL\rangle$, $|LR\rangle$, $|RL\rangle$ and $|RR\rangle$ for helicity eigenstate at given momenta, which will be known from context; in the same basis, the maximally entangled Bell states will also be defined as $|\phi^{\mp}\rangle = (|LL\rangle\mp |RR\rangle)/\sqrt2$ and 
$|\psi^{\mp}\rangle = (|LR\rangle\mp |RL\rangle)/\sqrt2$.

%%%%%%%%%%%%%%%%%%%%%%%%%%%%%%%%%%%%%
\subsection{Evaluation of two-qubit entanglement and entropy}

A separable state is one which can be created by local operations and classical communication (LOCC) and can therefore be written as \begin{equation}
    \rho_{sep} = \sum_j p_j \rho_{Aj} \otimes \rho_{Bj} \; ,
\end{equation} where $\rho_{Aj}$ and $\rho_{Bj}$ are density matrices for the reduced states for systems A and B, respectively. 
An entangled state is one that is not separable.
If one applies partially (that is, on only one of the systems) a positive (P) but not completely positive (CP) map $\Gamma$ on a separable state $\rho_{sep}$, one gets \begin{equation}
    (1 \otimes \Gamma) \rho_{sep} \geq 0 \, ,
\end{equation}since a density matrix must be positive semi-definite. 
Thus, a sufficient condition for $\rho$ to be entangled is \begin{equation}
    (1 \otimes \Gamma) \rho \ngeq 0 \; .
\end{equation} Transposition is one such map, so its partial action on the quantum state, the ``partial transposition'' \begin{equation}
    \rho^{T_B} = (1 \otimes T)\rho \label{parttr}
\end{equation} yields a general sufficient criterion for entanglement, called the Positivity of the Partial Transpose (PPT) or Peres-Horodecki criterion \cite{Peres1996, Horodecki1996}. For both 2-qubit and 1 qubit + 1 qutrit systems, the Peres-Horodecki criterion turns out to be both necessary and sufficient for entanglement.

The computation of the eigenvalues of the partially transposed density matrix is also helpful to quantify the amount of entanglement of the system, via the so-called negativity \begin{equation} \label{neg}
    \mathcal{N}(\rho) = \sum_i \frac{\abs{\lambda_i} - \lambda_i}{2} \; ,
\end{equation} where the $\lambda_i$ are the eigenvalues of the partially transposed density matrix. The logarithmic negativity, defined as \begin{equation} \label{lneg}
    E_\mathcal{N}(\rho) = \log_2(2 \mathcal{N} + 1) \; ,
\end{equation} is also a useful mathematical object as it constitutes an upper bound to the operationally defined distillable entanglement \cite{Vidal2002,Plenio2005}.
Such an entanglement monotone (i.e., a quantity that decreases under LOCC) 
is a proper measure of entanglement, which can be systematically evaluated for all two-qubit states.

Another relevant, information related quantity in this context 
is the von Neumann entropy, defined as
\begin{equation} \label{VNS}
    S(\rho) = - \sum_j \nu_j \ln(\nu_j)
\end{equation} where the $\nu_j$ are the eigenvalues of the density matrix. The von Neumann entropy quantifies the \textit{purity} of a state (or, conversely, the noise that affects it); when $S(\rho) = 0$ the state is pure (i.e., a one-dimensional projector or, more commonly put, a `wave-function'), whilst when $S(\rho)=\ln(4)$ the state is maximally mixed. 
It is worth noting that, although the scattering evolution we consider is of course unitary on the entire field, the final measurement leads to an output state which does not necessarily have the same entropy as the initial one. This occurs because the filtering of specific momenta selects field modes that are entangled with the remainder of the field, as a consequence of the global unitary interaction. Yet, such a filtering is not a projection on a one-dimensional subspace, since helicity is not detected (otherwise, the entanglement in the helicity basis would be compromised too), thus implying the possibility of a mixed filtered helicity state.
The inclusion of further modes and the study of multimode entanglement and correlation patterns in QED scattering at tree-level may well deserve further investigation in the future.

%%%%%%%%%%%%%%%%%%%%%%%%%%%%%%%%%%%%%%%%%%%%%%%%
\subsection{Method}

A test of the necessary and sufficient conditions for entanglement for any $2 \rightarrow 2$ scattering process in QED goes as follows: 

\paragraph{} Choose the initial state, i.e., set the parameters $p_{\lambda}$ in Eq.~(\ref{evol}). A common choice, which is the appropriate assumption when nothing is known about the initial helicity, is the maximally mixed initial state with $p_\lambda=1/4$ for $\lambda=1,\ldots,4$.

\paragraph{} Evaluate Feynman diagrams to the desired order and so obtain the quantities $\langle q,\eta|S|p,\lambda\rangle$ in Eq.~(\ref{rhozzo}). Our evaluation will be done at the lowest order, for diagrams with no loops (i.e., at tree-level).

\paragraph{} Use Eq.~(\ref{rhozzo}) to determine the output helicity density matrix elements $\rho_{\eta,\eta'}$.

\paragraph{} Derive the partially transposed output density matrix and evaluate its eigenvalues. If any of them is negative, then the state is entangled, otherwise it is separable. 

\paragraph{} Employ Eqs.~(\ref{neg}), (\ref{lneg}) and (\ref{VNS}) to evaluate the logarithmic negativity and von Neumann entropy of the final state.

The explicit expressions for the density matrix elements would unfortunately be extremely cumbersome and not very illuminating, so we shall not report them here although, as noted above, they can be retrieved through 
Eq.~(\ref{rhozzo}).

%%%%%%%%%%%%%%%%%%%%%%%%%%%%%%%%%%%%%%
\subsection{Loop Contributions}

Of course, the state and hence the entanglement determined through the methodology above are only accurate to a given perturbative order. 
In this paper, where only tree-level diagrams are considered, this is to the order $\alpha^2$ 
in terms of the fine structure constant $\alpha$.
Investigating the potential to create entanglement at tree-level is still an interesting question {\em per se}, although it may not necessarily give the full picture of quantum correlations in the outgoing state.

We may nevertheless apply the argument that higher order corrections to the density matrix, and hence to its partially transposed eigenvalues, are of order $\alpha^3$, up to factors which will be of order unity. Therefore, when the smallest partially transposed eigenvalue is such that $\abs{\min \{\lambda_i(p,\theta)\}} \gg \alpha^{3}$, one heuristically expects the separability or entanglement of the final state to be preserved at higher order. When, instead,$ \abs{\min \{\lambda_i(p,\theta)\}}$ is comparable with $\alpha^3$ (say around $10^{-6}$ in natural units, allowing for a possible factor), then it is well possible that the inclusion of higher order terms may entangle or disentangle separable or entangled tree-level states; we shall then say that the system has switching potential in the corresponding region of parameter.

%%%%%%%%%%%%%%%%%%%%%%%%%%%%%%%%%%%%%%%%%%%
\section{Entanglement in QED Scattering processes}

Let us now analyse systematically the tree-level helicity entanglement generated in the whole gallery of $2 \rightarrow 2$ scatterings in QED. 

%%%%%%%%%%%%%%%%%%%%%%%%%%%%%%%%%%%%%%%%%
\subsection{Møller Scattering} \label{eeee}

\begin{figure}[b!]
    \centering
    \subfloat[\centering t-channel]{{\begin{tikzpicture}
      \begin{feynman}[scale=0.75,transform shape]
        \vertex (gtop);
        \vertex [above left=of gtop] (etopleft) {\(e^- (s_1, p_1)\)};
        \vertex [below=of gtop] (gbot);
        \vertex [above right=of gtop] (etopright) {\(e^- (r_1, q_1)\)};
        \vertex [below left=of gbot] (ebotleft) {\(e^- (s_2, p_2)\)};
        \vertex [below right=of gbot] (ebotright) {\(e^- (r_2, q_2)\)};
    
        \diagram* {
            (etopleft) -- [fermion] (gtop),
            (gtop) -- [fermion] (etopright),
            (gtop) -- [photon, edge label'=\(\gamma\)] (gbot),
            (ebotleft) -- [fermion] (gbot),
            (gbot) -- [fermion] (ebotright),
        };
      \end{feynman}
    \end{tikzpicture}}}%
    \qquad
    \subfloat[\centering u-channel]{{\begin{tikzpicture}
      \begin{feynman}[scale=0.75,transform shape]
        \vertex (gtop);
        \vertex [above left=of gtop] (etopleft) {\(e^- (s_1, p_1)\)};
        \vertex [below=of gtop] (gbot);
        \vertex [above right=of gtop] (etopright) {\(e^- (r_1, q_1)\)};
        \vertex [below left=of gbot] (ebotleft) {\(e^- (s_2, p_2)\)};
        \vertex [below right=of gbot] (ebotright) {\(e^- (r_2, q_2)\)};
    
        \diagram* {
            (etopleft) -- [fermion] (gtop),
            (gbot) -- [fermion] (etopright),
            (gtop) -- [photon, edge label'=\(\gamma\)] (gbot),
            (ebotleft) -- [fermion] (gbot),
            (gtop) -- [fermion] (ebotright),
        };
      \end{feynman}
    \end{tikzpicture}}}
    
    \caption{Feynman diagrams of the t and u channels of a Møller scattering process ($e^- e^- \rightarrow e^- e^-$).}
    
    \label{fig:Feyneeee}%
\end{figure}
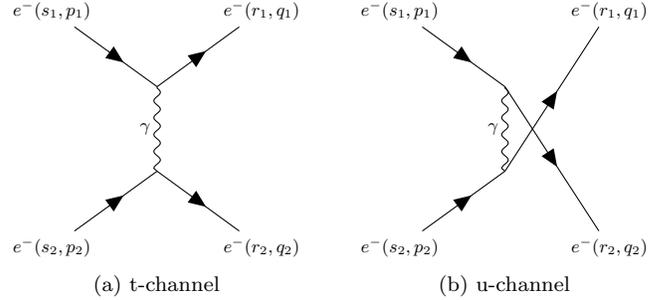
We first consider Møller scattering, that is, the process $e^- e^- \rightarrow e^- e^-$, for initially unpolarised particles ($p_\lambda=1/4$ $\forall \lambda$). The scattering amplitudes can be deduced from the Feynman diagrams describing this process shown in Fig.~\ref{fig:Feyneeee}, with

\begin{eqnarray}
    i \mathcal{M}_t &= \Bar{u}(r_1,q_1)(-ie\gamma^\mu)u(s_1,p_1) \frac{-i g_{\mu \nu}}{(p_1 - q_1)^2} \nonumber \\ &\times \Bar{u}(r_2, q_2) (-ie\gamma^\nu) u(s_2,p_2) \, , \\
    i \mathcal{M}_u &= -\Bar{u}(r_1,q_1)(-ie\gamma^\mu)u(s_2,p_2) \frac{-i g_{\mu \nu}}{(p_2 - q_1)^2} \nonumber \\ &\times \Bar{u}(r_2, q_2) (-ie\gamma^\nu) u(s_1,p_1) \, .
\end{eqnarray}

We can then expand the total scattering amplitude $\mathcal{M} = \mathcal{M}_t + \mathcal{M}_u$ in the helicity basis by setting each of the $s_1$, $s_2$, $r_1$ and $r_2$ to $R$ and $L$ (right- and left-handed, respectively) so as to get $\mathcal{M}_{\ket{RR} \rightarrow \ket{RR}}$, $\mathcal{M}_{\ket{RR} \rightarrow \ket{RL}}$, etc. This yields the $4 \times 4$ scattering matrix, which we can use with Eq.~(\ref{rhozzo}) to obtain $\rho_{out}$.

\begin{figure}[b!]
    \centering
    \includegraphics[scale=0.4]{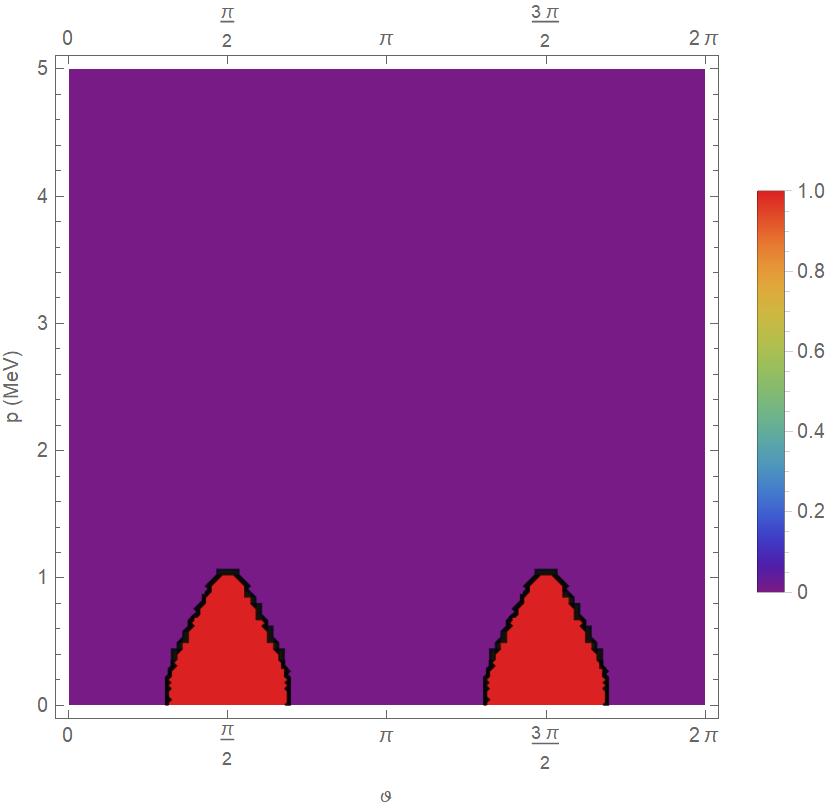}
    \caption{The red regions in this plot correspond to the values of $p$ and $\theta$ for which the final state is entangled.}
    \label{fig:eeeeMAP}
\end{figure}

In this case, we can compute the eigenvalues of the partially transposed density matrix analytically, and we find that, as one should expect, they are invariant under $\theta \rightarrow \theta + \pi$. It can be shown that three of these eigenvalues (that we call $\lambda_1$, $\lambda_3$ and $\lambda_4$) are positive $\forall p \in \mathbb{R}^+, \theta \in [0,2\pi[$. However, we have 
\begin{widetext}
\begin{equation}
    \lambda_2 < 0 \Leftrightarrow \cos (2 \theta )<-\frac{1}{3} \text{ \& } p<2 m_e \sqrt{\frac{\sqrt{(\cos (2 \theta
   )-9) (3 \cos (2 \theta )+1)} | \sin (\theta )| -6 \cos (2 \theta )-2}{28
   \cos (2 \theta )+\cos (4 \theta )+35}} \, .
\end{equation}
\end{widetext}
Thus, the Peres-Horodecki criterion provides us with a simple, necessary and sufficient condition to determine outgoing entanglement in terms of the COM momentum and scattering angle, which is depicted in Fig.~\ref{fig:eeeeMAP}. We see that the system is entangled for values of initial momenta in the COM frame of around the electron mass scale, and for scattering angles around $\theta = (n + \frac{1}{2}) \pi, n \in \mathbb{N}$, where the first inequality above is met. In particular, no entanglement is generated in the high energy limit.

In fact, the largest initial COM momentum for which the system is entangled is when $\theta= (n + \frac{1}{2}) \pi, n \in \mathbb{N}$, in which case \begin{equation}
    \lambda_2\big(p,\frac{\pi}{2}\big) < 0 \Leftrightarrow p < \sqrt{\sqrt5 +2}\, m_e \sim 1.05 \text{ MeV}
\end{equation}

\begin{figure}[b!]
    \centering
    \includegraphics[scale=0.4]{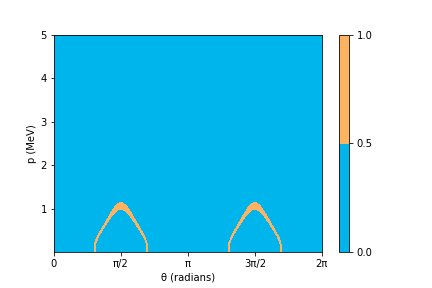}
    \caption{Regions where $\min\{\lambda_i(p,\theta)\}\le \alpha^3$ for Møller scattering with unpolarised input.}
    \label{fig:eeeeSWAP}
\end{figure}

The contribution of loops can be assessed heuristically, by highlighting regions in the $(p,\theta)$ space where scattering amplitudes of 1-loops may switch a product state to an entangled state and \textit{vice-versa}, i.e., the regions where the absolute value of the smallest partially transposed eigenvalue at tree-level is smaller than $\alpha^3$, which are shown in Fig.~\ref{fig:eeeeSWAP}.
We see that there is a possibility that such diagrams slightly expand or contract the main entanglement regions, but that there is no other isolated region in phase space where the system may become entangled. This also lends us a concrete idea of the accuracy of the tree-level approximation in evaluating entanglement.

\begin{figure}[t!]
    \centering
    {{\includegraphics[scale=0.4]{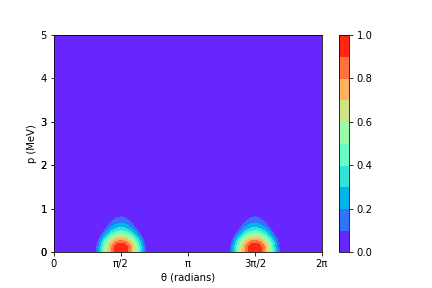} }}%
    \caption{Plot of the logarithmic negativity as a function of $p$ and $\theta$ for Møller scattering.}%
    \label{fig:eeeeNegs}
\end{figure}
The outgoing entanglement may be quantified through the logarithmic negativity, shown in Fig.~\ref{fig:eeeeNegs}.
Here, we see that the system is very entangled for very small momenta around $\theta = (n + \frac{1}{2}) \pi, n \in \mathbb{N}$, and that the negativity decreases with $p$ and as we move away from a scattering angle of $\theta = (n + \frac{1}{2}) \pi$.
\begin{figure}[t!]
    \centering
    \includegraphics[scale=0.4]{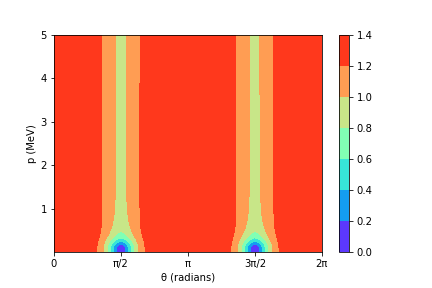}
    \caption{von Neumann Entropy of the system as a function of $p$ and $\theta$ for Møller scattering}
    \label{fig:eeeeVNS}
\end{figure}
The regions of maximal entanglement must necessarily be populated by pure states, which is indeed confirmed by a direct analysis of the von Neumann entropy of the system, shown in Fig.~\ref{fig:eeeeVNS}. 
We see that the system is maximally mixed around $\theta = n\pi, n \in \mathbb{N}$, whilst it is pure for small momenta around $\theta = (n + \frac{1}{2}) \pi, n \in \mathbb{N}$. 

It is very interesting to analyse in detail the mechanism whereby maximal entanglement emerges in this system. 
If the initial state is purified by filtering it into $|LL\rangle$ or $|RR\rangle$, then at $\theta=\pi/2$ and in the soft limit $p\rightarrow 0+$, one gets the maximally entangled output $|\phi^-\rangle$; if, instead, the initial state is $|LR\rangle$ or $|RL\rangle$, then one ends up with the orthogonal maximally entangled output $|\psi^-\rangle$. 
The dynamical generation of such entanglement 
was already discussed in  \cite{Cervera-Lierta2017}, and is due to the fact that, for $\theta=\pi/2$, the Mandelstam variables $t$ and $u$, driving the two channels pertaining to the process's two Feynman diagrams are equal, and as a consequence the interference between the two diagrams results into a balanced superposition.
If the evolution were linear, one would then expect an incoherent mixture of such two orthogonal states in the case of a completely unpolarised initial state. However, the measurement process is nonlinear in the input state, and in fact the effect of measurement normalisation in this instance is to completely suppress the $|\psi^-\rangle$ contribution to the mixture: operationally this is reflected by the fact that, at a perpendicular scattering angle and in the limit $p\rightarrow0$, all output detected particles come from the input branch with the same polarisation (i.e., the probability of detecting particles with different polarisations vanishes).
In fact, in the non-relativistic limit ($p << m_e$), the process's differential 
cross section is given by
$\frac{d\sigma}{d\Omega} = \frac{m_e^2 \alpha^2}{4 p^4 \sin^4(\theta)}(1 + 3\cos^2(\theta))$, with divergences in the soft scattering limit $p\rightarrow0$ and for the (non-entangled) output at $\theta=0$ and $\theta=\pi$.

%%%%%%%%%%%%%%%%%%%%%%%%%%%%%%%%%%%%%%%%%%%%%%%%%%
\subsection{Electron-Positron to Muon-Antimuon Pair} \label{eemm}

\begin{figure}[h!]
    \centering
    \begin{tikzpicture}
      \begin{feynman}[scale=0.75,transform shape]
        \vertex (gleft);
        \vertex [above left=of gleft] (e) {\(e^- (s_1, p_1)\)};
        \vertex [below left=of gleft] (p) {\(e^+ (s_2, p_2)\)};
        \vertex [right=of gleft] (gright);
        \vertex [above right=of gright] (mum) {\(\mu^- (r_1, q_1)\)};
        \vertex [below right=of gright] (mup) {\(\mu^+ (r_2, q_2)\)};

        \diagram* {
            (e) -- [fermion] (gleft),
            (p) -- [anti fermion] (gleft),
            (gleft) -- [photon, edge label'=\(\gamma\)] (gright),
            (gright) -- [fermion] (mum),
            (gright) -- [anti fermion] (mup),
        };
      \end{feynman}
    \end{tikzpicture}
    \caption{Feynman diagram of an electron-positron annihilation to a muon-antimuon pair ($e^- e^+ \rightarrow \mu^- \mu^+$).}
    \label{fig:Feyneemm}
\end{figure}
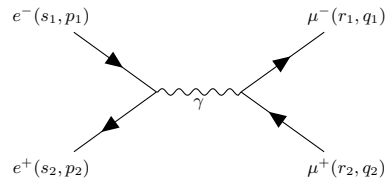

Let us now examine muon-antimuon pair creation from the scattering of an electron-positron pair, that is, the process $e^- e^+ \rightarrow \mu^- \mu^+$. At tree-level, this process is described by a single Feynman diagram, shown in Fig.~\ref{fig:Feyneemm}, with scattering amplitude  \begin{eqnarray}
    i \mathcal{M} &= \Bar{v}(s_2,p_2)(-ie\gamma^\mu)u(s_1,p_1) \frac{-i g_{\mu \nu}}{(p_1 + p_2)^2}  \nonumber \\ &\times \Bar{u}(r_1, q_1) (-ie\gamma^\nu) v(r_2,q_2) \, . \label{eemmM}
\end{eqnarray}

\begin{figure}[b!]
    \centering
    \includegraphics[scale=0.34]{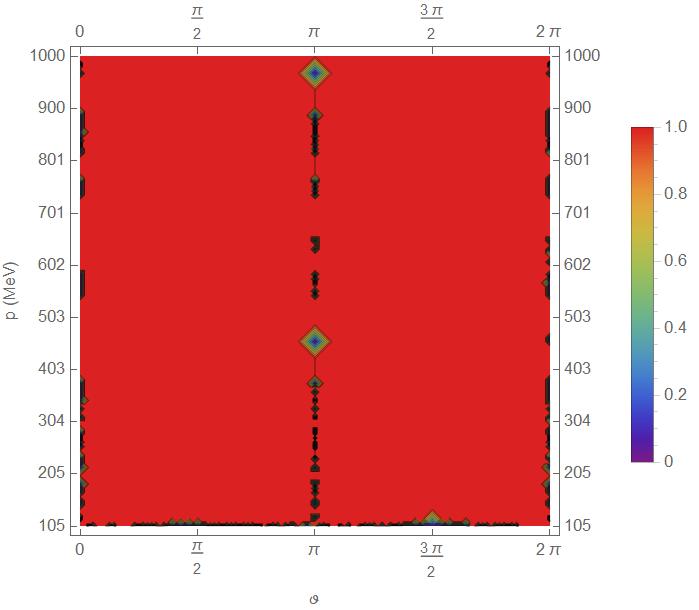}
    \caption{The red region in this plot corresponds to the values of $p$ and $\theta$ for which the final state is entangled.}
    \label{fig:eemmMAP}
\end{figure}

The eigenvalues of the partially transposed output state are also all invariant under $\theta\rightarrow\theta+\pi$ and may be determined analytically for an unpolarised input. Only one of them (it is a general feature of the partial transposition of two-qubit density matrices to allow for a single negative eigenvalue), which we refer to as $\lambda_1$, is negative for certain values of $p$ and $\theta$. Quite remarkably, as can be seen in Fig.~\ref{fig:eemmMAP}, the pair of output muons is entangled for most of the parameter space.

\begin{figure}[t!]
    \centering
    \subfloat[\centering\label{fig:eemmLNEG}]{{\includegraphics[height=3cm]{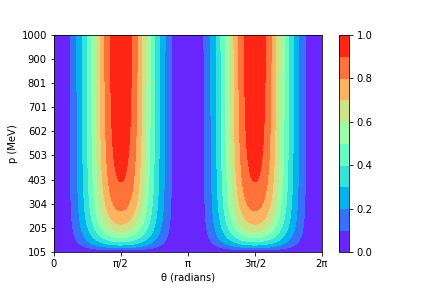}}}%
    \quad
    \subfloat[\centering\label{fig:eemmVNS}]{{\includegraphics[height=3cm]{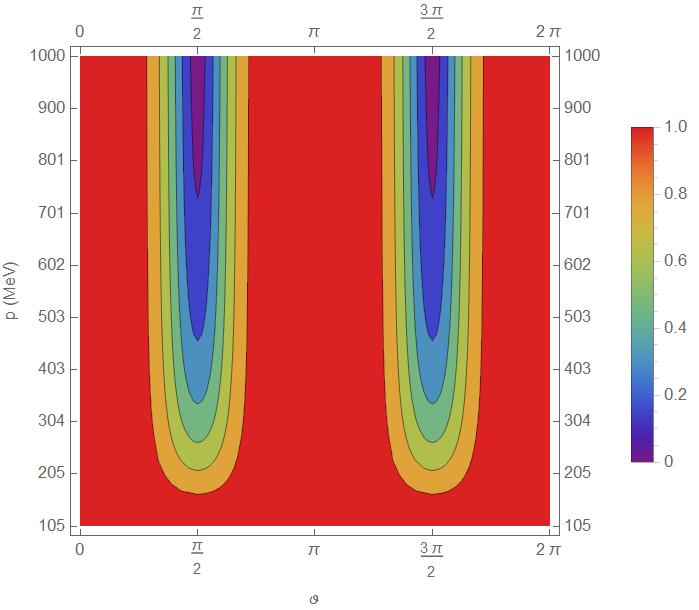}}}%
    \caption{Plot of the logarithmic negativity (a) and von Neumann entropy (b) as functions of $p$ and $\theta$ for the scattering $e^- e^+ \rightarrow \mu^- \mu^+$.}
\end{figure}

We see that the system is entangled for all values of $p$ and $\theta$ except in the vicinity of $\theta = n\pi, n \in \mathbb{N}$, as well as around the minimum of the energy $p = \sqrt{m_\mu^2 - m_e^2}$.
This analysis is reflected in the logarithmic negativity, plotted in Fig.~\ref{fig:eemmLNEG} : the system is least entangled around $p = \sqrt{m_{\mu}^2 - m_e^2}$ and around $\theta = n \pi, n \in \mathbb{N}$, whilst it is maximally entangled around $\theta = (n + \frac{1}{2})\pi, n \in \mathbb{N}$ for very large momenta. Accordingly, the state is mixed around $p = \sqrt{m_{\mu}^2 - m_e^2}$ and $\theta = n \pi, n \in \mathbb{N}$, whilst it is pure for $\theta = (n+\frac{1}{2}) \pi, n \in \mathbb{N}$ at very large energies, as shown in Fig.~\ref{fig:eemmVNS}. 

At high energies at $\theta = \pi/2$, the final state is the pure Bell state $|\psi^-\rangle$. Conversely to what was previously seen in the case of Møller scattering, this is an instance where outgoing particles coming from different initial helicities dominate the output in the high energy limit, which is thus purified into a Bell state. The $s$-channel generation of such maximal entanglement was already discussed in \cite{Cervera-Lierta2017} (where it is argued that the virtual photon cannot distinguish, so to speak, between the helicities $\ket{LR}$ and $\ket{RL}$, and thus carries an equal proportion of the currents, which favours maximal entanglement generation). The filtering of initial states with equal helicities would instead result in a $|\psi^{+}\rangle$ Bell state at $\theta=\pi/2$ in the ultra-relativistic limit. In the collinear ($\theta=0$) direction, equal initial helicities give rise to the $|\phi^-\rangle$ Bell state but different initial helicities stay pure and separable, which results into a mixed state. 

These entanglement features could be weighted against the behaviour of the differential cross section for an unpolarised beam, which behaves like $[1+(\cos\theta)^2]$ at high energies: quite significantly, the probability of detecting outgoing particles is in this case minimum around the region where maximal entanglement is generated.

%%%%%%%%%%%%%%%%%%%%%%%%%%%%%%%%%%%%%%%%%%%%%
\subsection{Electron-Positron Annihilation} \label{EPA}

\begin{figure}[h!]
    \centering
    \subfloat[\centering t-channel]{{\begin{tikzpicture}
      \begin{feynman}[scale=0.75,transform shape]
        \vertex (etop);
        \vertex [above left=of etop] (e) {\(e^-(s_1,p_1)\)};
        \vertex [below=of etop] (ebot);
        \vertex [above right=of etop] (gtop) {\(\gamma(\lambda_1,q_1)\)};
        \vertex [below left=of ebot] (p) {\(e^+(s_2,p_2)\)};
        \vertex [below right=of ebot] (gbot) {\(\gamma(\lambda_2,q_2)\)};
    
        \diagram* {
            (e) -- [fermion] (etop),
            (etop) -- [fermion] (ebot),
            (etop) -- [photon] (gtop),
            (ebot) -- [fermion] (p),
            (ebot) -- [photon] (gbot),
        };
      \end{feynman}
    \end{tikzpicture}}}%
    \qquad
    \subfloat[\centering u-channel]{{\begin{tikzpicture}
      \begin{feynman}[scale=0.75,transform shape]
        \vertex (etop);
        \vertex [above left=of etop] (e) {\(e^-(s_1,p_1)\)};
        \vertex [below=of etop] (ebot);
        \vertex [above right=of etop] (gtop) {\(\gamma(\lambda_1,q_1)\)};
        \vertex [below left=of ebot] (p) {\(e^+(s_2,p_2)\)};
        \vertex [below right=of ebot] (gbot) {\(\gamma(\lambda_2,q_2)\)};
    
        \diagram* {
            (e) -- [fermion] (etop),
            (etop) -- [fermion] (ebot),
            (etop) -- [photon] (gbot),
            (ebot) -- [fermion] (p),
            (ebot) -- [photon] (gtop),
        };
      \end{feynman}
    \end{tikzpicture}}}
    
    \caption{Feynman diagrams of the t and u channels of an electron-positron annihilation ($e^- e^+ \rightarrow \gamma \gamma$).}
    
    \label{fig:Feyneegg}%
\end{figure}
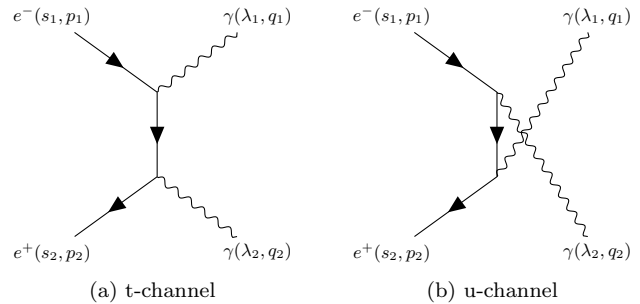

We now consider an electron-positron annihilation process into two photons, i.e. the process $e^- e^+ \rightarrow \gamma \gamma$. From the Feynman diagrams shown in Fig.~\ref{fig:Feyneegg} describing this process, the scattering matrices can be read as \begin{eqnarray}
    i \mathcal{M}_t &= -ie^2 \epsilon_\mu^*(\lambda_1,q_1) \epsilon_\nu^*(\lambda_2,q_2) \nonumber \\ &\times \Bar{v}(s_2,p_2) \gamma^\mu \frac{i}{\slashed{t} - m} \gamma^\nu u(s_1,p_1) \, , \label{eeggM1} \\
    i \mathcal{M}_u &= -ie^2 \epsilon_\mu^*(\lambda_1,q_1) \epsilon_\nu^*(\lambda_2,q_2) \nonumber \\ &\times \Bar{v}(s_2,p_2) \gamma^\mu \frac{i}{\slashed{u} - m} \gamma^\nu u(s_1,p_1) \, , \label{eeggM2}
\end{eqnarray}
where $\mathcal{M}_t$ and $\mathcal{M}_u$ are the scattering amplitudes corresponding to the t and u channels, respectively.

\begin{figure}[t!]
    \centering
    \subfloat[\centering \label{fig:eeggMAP}]{{\includegraphics[height=3cm]{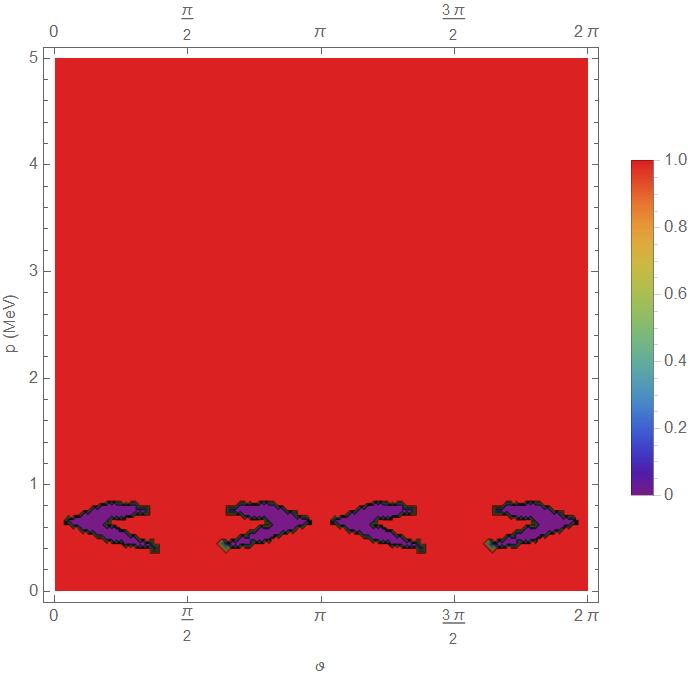}}}%
    \quad
    \subfloat[\centering  \label{fig:eeggLNEG}]{{\includegraphics[height=3cm]{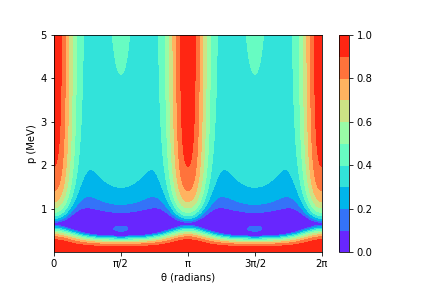}}}%
    \caption{Plot of the PPT criterion (the red region in this plot corresponds to the region for which the final state is entangled) and logarithmic negativity as functions of $p$ and $\theta$ for electron-positron annihilation.}
\end{figure} 

\begin{figure}[b!]
    \centering
    \includegraphics[scale=0.26]{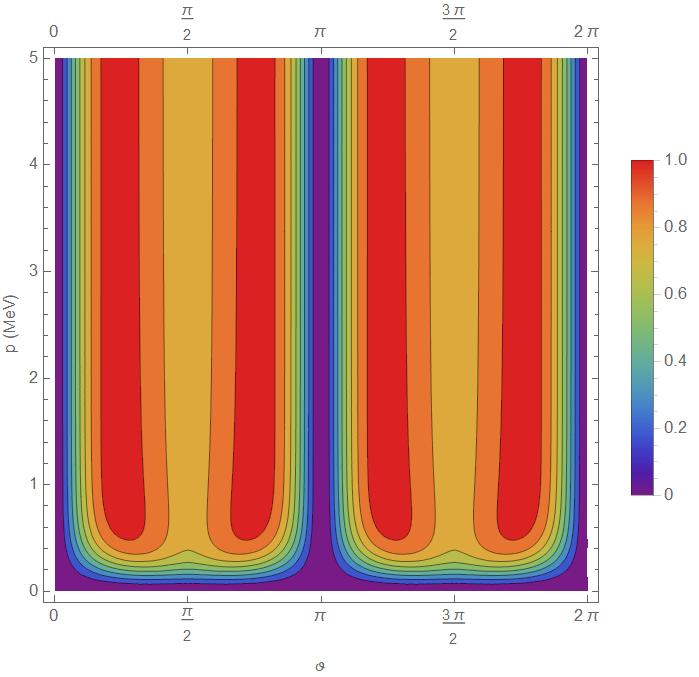}
    \caption{von Neumann Entropy of the system as a function of $p$ and $\theta$ for electron-positron annihilation.}
    \label{fig:eeggVNS}
\end{figure}

As before, we can compute the scattering matrix expanded in the helicity basis, and hence the resulting final density matrix for an unpolarised input and its partial-transposed and eigenvalues, periodic in $\theta \rightarrow \theta + \pi$.

Two eigenvalues contribute to the entanglement of the system by the PPT criterion. The whole of phase space is entangled except in regions that we call the ``wing domains", as can be seen in Fig.~\ref{fig:eeggMAP}. These can be shown numerically to span between $p=\frac{m_e}{\sqrt{2}} \approx 0.36 \text{ MeV}$ and $p \approx 0.9 \text{MeV}$ and are symmetrical around $\theta = \frac{n}{2} \pi, n \in \mathbb{N}$.

By looking at the logarithmic negativity, plotted in Fig.~\ref{fig:eeggLNEG}, we see that the system is most entangled for small momenta at all angles $\theta$ as well as for high momenta for $\theta = n \pi, n \in \mathbb{N}$. On the other hand, around $\theta = \pi/2$ at large energies, the system is almost in a product state and is barely entangled.

We can also look at the von Neumann entropy of the final system, shown in Fig.~\ref{fig:eeggVNS}, and we see that the system is pure at low energies and around $\theta = n\pi, n \in \mathbb{N}$, whilst it is maximally mixed in the wing domains.

At low energies, the same helicity input branch dominates the output and imposes the maximally entangled state  $\ket{\phi^+}$, whilst in the ultra-relativistic limit at $\theta =0$ the maximally entangled state $\ket{\phi^-}$ emerges.

%%%%%%%%%%%%%%%%%%%%%%%%%%%%%%%%%%%%%
\subsection{Bhabha Scattering} \label{epep}

\begin{figure}[h!]
    \centering
    \subfloat[\centering s-channel]{{\begin{tikzpicture}
      \begin{feynman}[scale=0.65,transform shape]
        \vertex (gleft);
        \vertex [above left=of gleft] (eleft) {\(e^-(s_1, p_1)\)};
        \vertex [below left=of gleft] (pleft) {\(e^+(s_2,p_2)\)};
        \vertex [right=of gleft] (gright);
        \vertex [above right=of gright] (eright) {\(e^-(r_1,q_1)\)};
        \vertex [below right=of gright] (pright) {\(e^+(r_2,q_2)\)};

        \diagram* {
            (eleft) -- [fermion] (gleft),
            (pleft) -- [anti fermion] (gleft),
            (gleft) -- [photon, edge label'=\(\gamma\)] (gright),
            (gright) -- [fermion] (eright),
            (gright) -- [anti fermion] (pright),
        };
      \end{feynman}
    \end{tikzpicture}}}%
    \qquad
    \subfloat[\centering t-channel]{{\begin{tikzpicture}
      \begin{feynman}[scale=0.65,transform shape]
        \vertex (gtop);
        \vertex [above left=of gtop] (eleft) {\(e^-(s_1,p_1)\)};
        \vertex [below=of gtop] (gbot);
        \vertex [above right=of gtop] (eright) {\(e^-(r_1,q_1)\)};
        \vertex [below left=of gbot] (pleft) {\(e^+(s_2,p_2)\)};
        \vertex [below right=of gbot] (pright) {\(e^+(r_2,q_2)\)};
    
        \diagram* {
            (eleft) -- [fermion] (gtop),
            (gtop) -- [fermion] (eright),
            (gtop) -- [photon, edge label'=\(\gamma\)] (gbot),
            (pleft) -- [anti fermion] (gbot),
            (gbot) -- [anti fermion] (pright),
        };
      \end{feynman}
    \end{tikzpicture}}}
    
    \caption{Feynman diagrams of the s and t channels of a Bhabha scattering ($e^- e^+ \rightarrow e^- e^+$).}
    
    \label{fig:Feynepep}%
\end{figure}
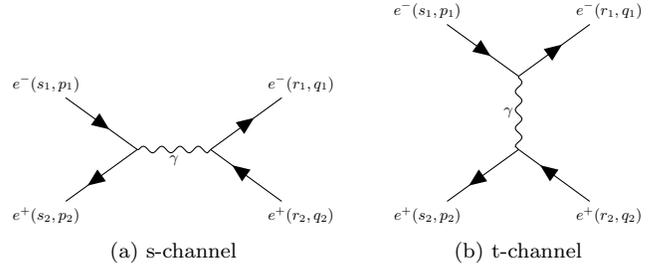
Bhabha scattering corresponds to an electron-positron pair scattering, i.e. $e^- e^+ \rightarrow e^- e^+$. From the Feynman diagrams shown in Fig.~\ref{fig:Feynepep} describing this process, the scattering amplitudes can be read as \begin{eqnarray}
    i \mathcal{M}_s &= \Bar{v}(s_2,p_2)(-ie\gamma^\mu)u(s_1,p_1) \frac{-i g_{\mu \nu}}{(p_1 + p_2)^2} \nonumber \\ &\times \Bar{u}(r_1, q_1) (-ie\gamma^\nu) v(r_2,q_2) \, , \label{epepM1} \\
    i \mathcal{M}_t &= -\Bar{v}(s_2,p_2)(-ie\gamma^\mu) v(r_2,q_2) \frac{-i g_{\mu \nu}}{(p_2 - q_1)^2} \nonumber \\ &\times \Bar{u}(r_1, q_1) (-ie\gamma^\nu) u(s_1,p_1) \, , \label{epepM2}
\end{eqnarray}
where $\mathcal{M}_s$ and $\mathcal{M}_t$ correspond to the $s$ and $t$ channels, respectively.

Again, the partially transposed eigenvalues of the density matrix derived from these amplitudes for unpolarised input can be evaluated analytically, and one can show that three of them are always positive whilst one, which we denote with $\lambda_1$, can be negative. At variance with the previous ones, this process is not invariant under $\theta \rightarrow \theta + \pi$ but only under $\theta\rightarrow -\theta$. 

\begin{figure}[t!]
    \centering
    \subfloat[\centering \label{fig:epepLNEG}]{{\includegraphics[height=3cm]{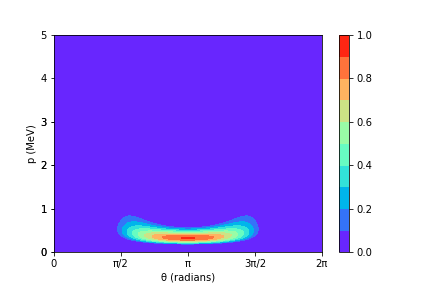}}}%
    \quad
    \subfloat[\centering \label{fig:epepVNS}]{{\includegraphics[height=3cm]{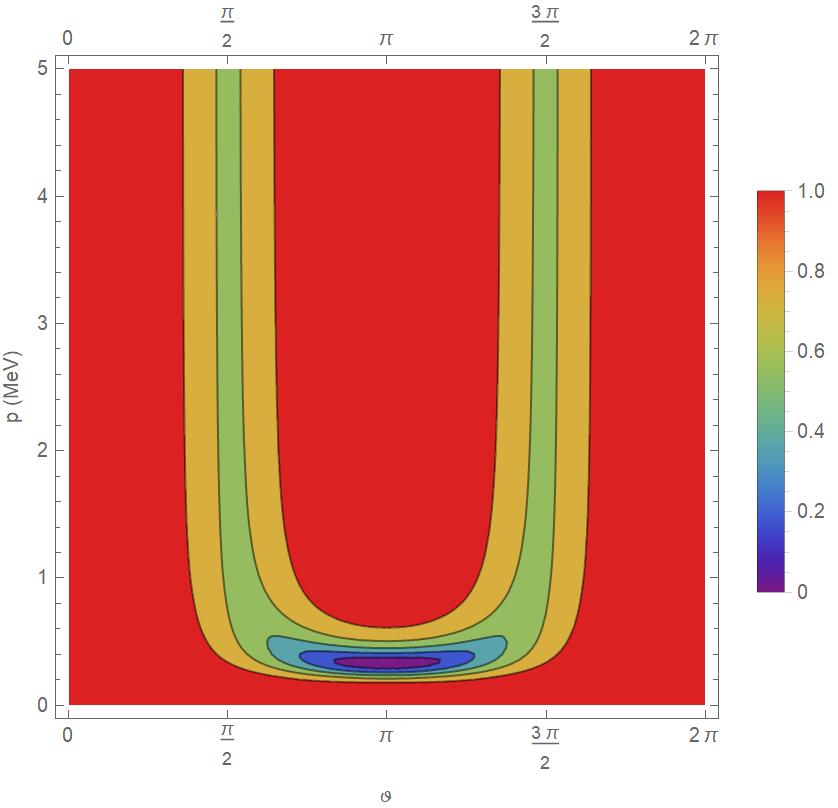}}}%
    \caption{Plot of the logarithmic negativity and von Neumann entropy as functions of $p$ and $\theta$ for Bhabha scattering.}
\end{figure}

The application of the Peres-Horodecki criterion is shown by plotting the logarithmic negativity in Fig.~\ref{fig:epepLNEG}. On this occasion, output  entanglement is only found in a fairly restricted region of parameters.
The study of the logarithmic negativity, backed by the von Neumann entropy in Fig.~\ref{fig:epepVNS}, allows us to visualise the point of maximal output entanglement at $\theta = \pi$ (with the particles bouncing back from each other) and $p \approx 0.32 \text{ MeV}$, 
where the output state is given by 
\begin{equation}
    \rho_{out} = 0.98 \ket{\phi^+}\bra{\phi^+} +              0.01 \ket{LR}\bra{LR} +
                 0.01 \ket{RL}\bra{RL} \, .
\end{equation}
Hence, maximal entanglement is very closely approximated at intermediate energies, comparable with the electron mass, in Bhabha scattering too, even for maximally mixed initial states. The $\ket{LR}$ and $\ket{RL}$ contributions to the state above come, respectively, from the $\ket{RL}$ and $\ket{LR}$ branches, for which this scattering acts as a helicity flip, and which are suppressed albeit not entirely. If these branches are filtered out by selecting initial particles with the same helicity, then one obtains a state which is, to all practical purposes (given the substantial noise one may anticipate in such processes) indistinguishable from the maximally entangled Bell state $\ket{\phi^+}$. It is worthwhile to note that the generation of such entanglement at intermediate energies occurs entirely through the $t$-channel, which dominates the process. 
This non-trivial finding does not feature in the analysis of \cite{Cervera-Lierta2017}, which focuses on the occurrence of exact maximal entanglement, especially in the low and high energy limit.

%\begin{figure}[b!]
%    \centering
%    \includegraphics[scale=0.4]{epep/epepEMAP.jpg}
%    \caption{The red region in this plot corresponds to the values of $p$ and $\theta$ for which the final state is entangled.}
%    \label{fig:epepMAP}
%\end{figure}

%%%%%%%%%%%%%%%%%%%%%%%%%%%%%%%%%%%%%%
\subsection{Electron-Muon Scattering} \label{emem}

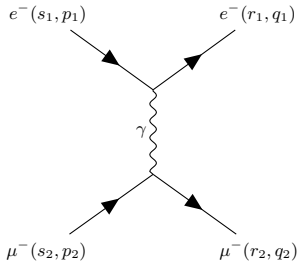
\begin{figure}[h!]
    \centering
    \begin{tikzpicture}
      \begin{feynman}[scale=0.75,transform shape]
        \vertex (gtop);
        \vertex [above left=of gtop] (eleft) {\(e^-(s_1,p_1)\)};
        \vertex [below=of gtop] (gbot);
        \vertex [above right=of gtop] (eright) {\(e^-(r_1,q_1)\)};
        \vertex [below left=of gbot] (muleft) {\(\mu^-(s_2,p_2)\)};
        \vertex [below right=of gbot] (muright) {\(\mu^-(r_2,q_2)\)};

        \diagram* {
            (eleft) -- [fermion] (gtop),
            (gtop) -- [fermion] (eright),
            (gtop) -- [photon, edge label'=\(\gamma\)] (gbot),
            (muleft) -- [fermion] (gbot),
            (gbot) -- [fermion] (muright),
        };
      \end{feynman}
    \end{tikzpicture}
    \caption{Feynman diagram of an electron-muon scattering process ($e^- \mu^- \rightarrow e^- \mu^-$)}
    \label{fig:Feynemem}
\end{figure}

Let us now analyse electron-muon scattering $e^- \mu^- \rightarrow e^- \mu^-$. From the Feynman diagram shown in Fig.~\ref{fig:Feynemem} describing this process, the scattering amplitude can be read as \begin{eqnarray}
    i \mathcal{M} &= \Bar{u}(r_2,q_2)(-ie\gamma^\mu)u(s_2,p_2) \frac{-i g_{\mu \nu}}{(p_1 - q_1)^2} \nonumber \\ &\times \Bar{u}(r_1, q_1) (-ie\gamma^\nu) u(s_1,p_1) \, . \label{ememM}
\end{eqnarray}

\begin{figure}[t!]
    \centering
    \subfloat[\centering ]{{\includegraphics[width=4cm]{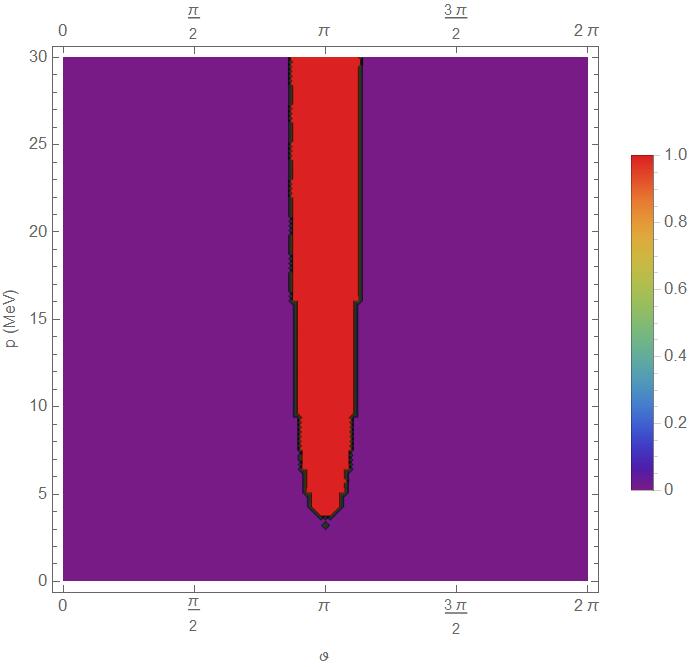} }}%
    \quad
    \subfloat[\centering]{{\includegraphics[width=4cm]{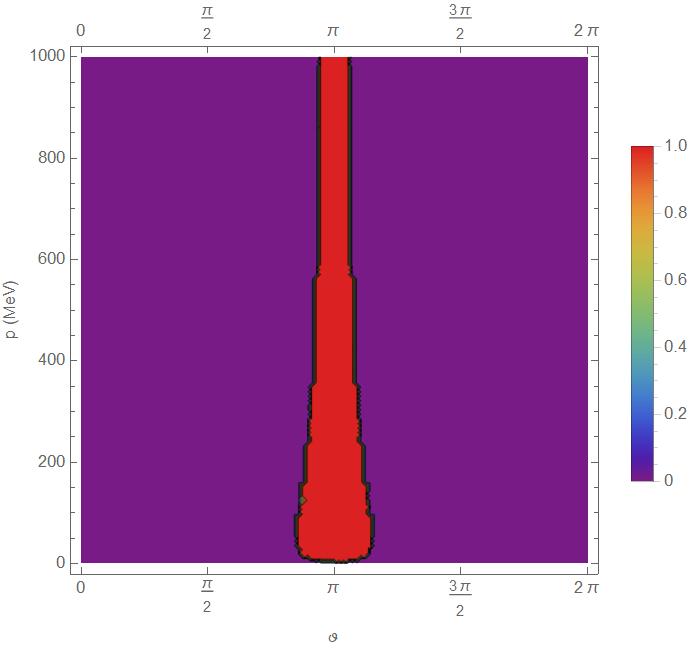}}}%
    \caption{The red regions correspond to the values of $p$ and $\theta$ for which the final state is entangled at low -- $<30$ MeV -- (a) and high -- $<1$ GeV -- (b) energies.}%
    \label{fig:ememMAP}
\end{figure}

From these, the only partially transposed eigenvalue that can be negative may be determined for unpolarised input to show that, by the PPT criterion, the output particles are entangled for \begin{equation}
    p > \frac{\sqrt{m_e m_\mu}}{2} \text{ and } \theta = \pi \pm \epsilon(p) \, ,
\end{equation} where $\epsilon(p)$ cannot be determined analytically but is small, increasing for $p \lesssim 50 \text{ MeV}$ but decreasing for large values of $p$. This is shown in Fig.~\ref{fig:ememMAP}.

\begin{figure}[b!]
    \centering
    \subfloat[\centering ]{{\includegraphics[width=4cm]{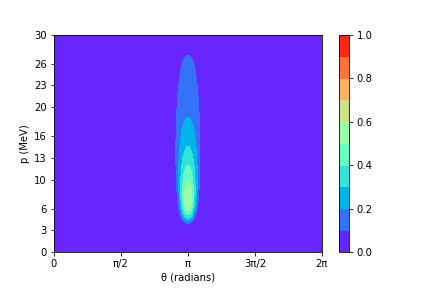}}}%
    \quad
    \subfloat[\centering ]{{\includegraphics[width=4cm]{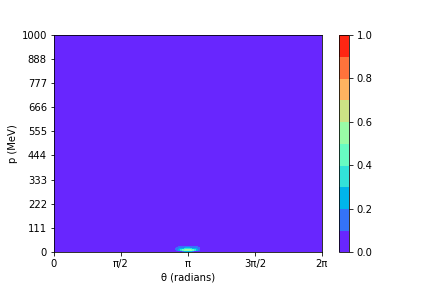}}}%
    \caption{Plots of the logarithmic negativity at low and high energies as a function of $p$ and $\theta$ for electron-muon scattering at low -- $<30$ MeV -- (a) and high -- $<1$ GeV -- (b) energies.}%
    \label{fig:ememLNeg}
\end{figure}
The logarithmic negativity is plotted in Fig.~\ref{fig:ememLNeg}, where one can see
that the system has its largest entanglement for $\theta = \pi$ and  $p = \sqrt{m_e m_\mu}$, whilst it is not very entangled for large momenta. For $p = \sqrt{m_e m_\mu}, \theta = \pi$, we have 
\begin{align}
    \rho_{out} & = \frac23 \ket{\phi^-}\bra{\phi^-} +\frac13 \frac{\mathbb{1}_{4\times4}}{4} \, .
\end{align}
Thus, we see that the system never reaches maximal entanglement, but approaches a noisy, depolarised Bell state for the identified values of $\theta$ and $p$.

\begin{figure}[t!]
    \centering
    \subfloat[\centering ]{{\includegraphics[width=4cm]{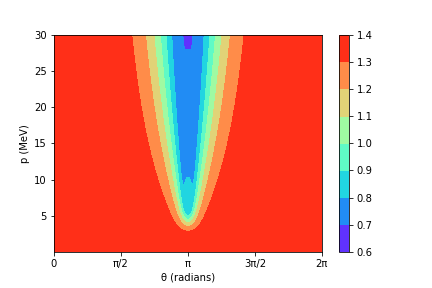}}}%
    \quad
    \subfloat[\centering ]{{\includegraphics[width=4cm]{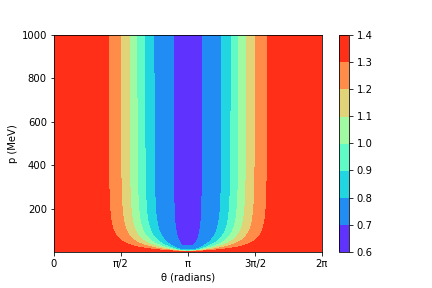}}}%
    \caption{Plots of the von Neumann Entropy at low  -- $<30$ MeV -- (a) and high -- $<1$ GeV -- (b) energies as a function of $p$ and $\theta$ for electron-muon scattering}%
    \label{fig:ememVNS}
\end{figure}

The von Neumann entropy of the system can be seen in Fig.~\ref{fig:ememVNS}, and one observes that the state is pure for $\theta = (2n + 1) \pi, n \in \mathbb{N}$ whilst it is maximally mixed when $\theta = 2n \pi, n \in \mathbb{N}$. In this case, at variance with the scattering processes examined so far, maximal purity does not correspond to maximal entanglement as these pure states are product states. 

%%%%%%%%%%%%%%%%%%%%%%%%%%%%%%%%%%%%%%
\subsection{Compton Scattering} \label{egeg}

% $e^- \gamma \rightarrow e^- \gamma$

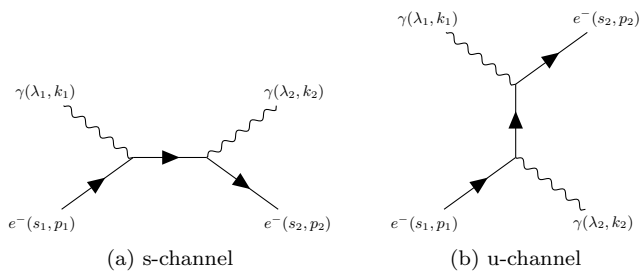
\begin{figure}[h!]
    \centering
    \subfloat[\centering s-channel]{\begin{tikzpicture}
      \begin{feynman}[scale=0.65,transform shape]
        \vertex (vleft);
        \vertex [above left=of vleft] (gleft) {\(\gamma (\lambda_1, k_1)\)};
        \vertex [below left=of vleft] (eleft) {\(e^- (s_1,p_1)\)};
        \vertex [right=of vleft] (vright);
        \vertex [above right=of vright] (gright) {\(\gamma  (\lambda_2, k_2)\)};
        \vertex [below right=of vright] (eright) {\(e^- (s_2,p_2)\)};

        \diagram* {
            (eleft) -- [fermion] (vleft),
            (gleft) -- [photon] (vleft),
            (vleft) -- [fermion] (vright),
            (vright) -- [fermion] (eright),
            (vright) -- [photon] (gright),
        };
      \end{feynman}
    \end{tikzpicture}}%
    \qquad
    \subfloat[\centering u-channel]{\begin{tikzpicture}
      \begin{feynman}[scale=0.65,transform shape]
        \vertex (vtop);
        \vertex [above left=of vtop] (gleft) {\(\gamma (\lambda_1,k_1)\)};
        \vertex [below=of vtop] (vbot);
        \vertex [below left=of vbot] (eleft) {\(e^- (s_1,p_1)\)};
        \vertex [above right=of vtop] (eright) {\(e^- (s_2,p_2)\)};
        \vertex [below right=of vbot] (gright) {\(\gamma (\lambda_2,k_2)\)};

        \diagram* {
            (eleft) -- [fermion] (vbot),
            (vbot) -- [fermion] (vtop),
            (vtop) -- [fermion] (eright),
            (gleft) -- [photon] (vtop),
            (vbot) -- [photon] (gright),
        };
      \end{feynman}
    \end{tikzpicture}}
    
    \caption{Feynman diagrams of the s and u channels for Compton scattering ($e^- \gamma \rightarrow e^- \gamma$).}
    
    \label{fig:Feynegeg}%
\end{figure}

Finally, we shall examine Compton scattering, that is, the process $e^- \gamma \rightarrow e^- \gamma$. From the Feynman diagrams shown in Fig.~\ref{fig:Feynegeg} describing this process, the scattering matrices can be read as \begin{align}
    i \mathcal{M}_s &= -i e^2 \Bar{u}(s_2,p_2) \gamma^\mu \epsilon_\mu^*(\lambda_2,k_2) \frac{\slashed{p} + \slashed{k} + m}{s^2 - m^2} \nonumber \\ &\gamma^\nu \epsilon_\nu(\lambda_1,k_1) u(s_1,p_1)\, , \label{egegM1} \\
    i \mathcal{M}_u &= -i e^2 \Bar{u}(s_2,p_2) \gamma^\nu \epsilon_\nu^*(\lambda_1,k_1) \frac{\slashed{p} - \slashed{k} + m}{u^2 - m^2} \nonumber \\ &\gamma^\mu \epsilon_\mu(\lambda_2,k_2) u(s_1,p_1) \, , \label{egegM2}
\end{align}
where $\mathcal{M}_s$ and $\mathcal{M}_u$ are the scattering amplitudes corresponding to the s and u channels, respectively.

As before, we can expand these in the helicity basis, time-evolve the initial state, apply partial transposition and impose the Peres-Horodecki criterion to determine the entanglement of the system, although in this case the final state for an unpolarised initial state is never entangled, regardless of the energy scale or the scattering angle.

\begin{figure}[t!]
    \centering
    \subfloat[\centering \label{fig:egegSWAP}]{{\includegraphics[height=3.1cm]{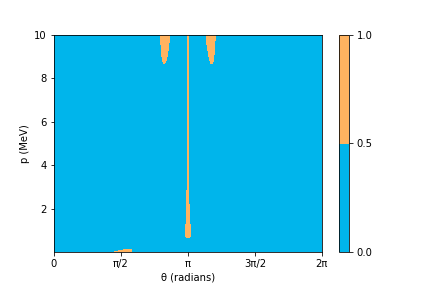}}}%
    \quad
    \subfloat[\centering \label{fig:egegVNS}]{{\includegraphics[scale=0.22]{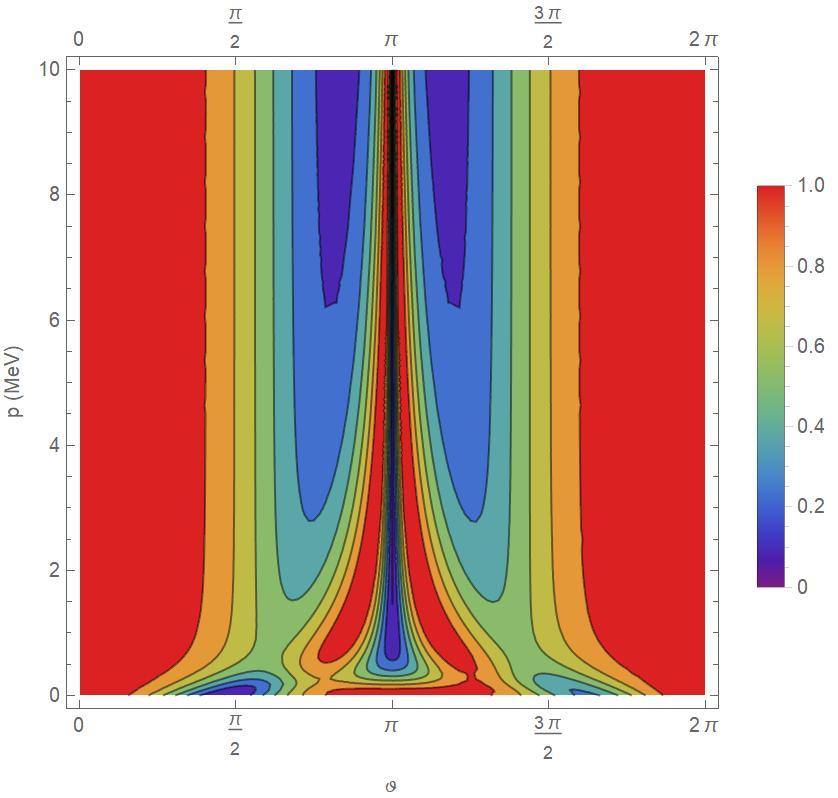}}}%
    \caption{(a) Regions where $\min\{\lambda_i(p,\theta)\}\le \alpha^3$ for Compton scattering with unpolarised input; all partially transposed eigenvalues at tree level are always positive in this case. (b) von Neumann entropy as a function of $p$ and $\theta$ for Compton scattering with unpolarised input.}%
\end{figure}

This is rather surprising and contrasts with other processes seen in QED. There is however the possibility for the system to be entanglable through loop diagrams, as can be seen in Fig.~\ref{fig:egegSWAP}, where parameter regions with partially transposed eigenvalues smaller than $\alpha^3$ are shown; note that such regions extend also beyond the infrared divergence that characterises the process at $\theta =\pi$.
The possibility of entanglement occurring at higher order is all the more likely given that, as shown by evaluating the von Neumann entropy plotted in Fig.~\ref{fig:egegVNS}, the states in those regions are very close to pure states. 

Even at tree-level, however, entanglement can actually be generated in Compton scattering, as we are about to see.

%%%%%%%%%%%%%%%%%%%%%%%%%%%%%%%%%%
\subsubsection{Werner state filtering}

\begin{figure}[t!]
    \centering
    \subfloat[\centering \label{fig:egegWernerEMAP}]{{\includegraphics[height=2.6cm]{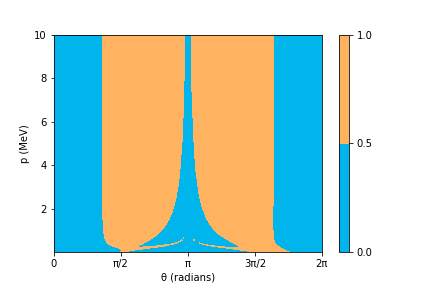}}}%
    \quad
    \subfloat[\centering \label{fig:egegWernerLNEG}]{{\includegraphics[scale=0.25]{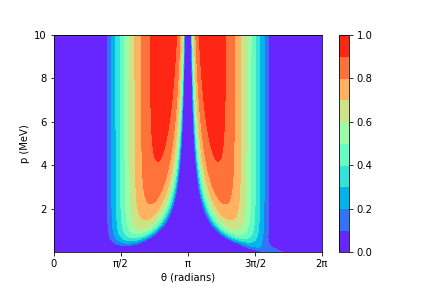}}}%
    \caption{(a) The orange regions in this plot correspond to the values of $p$ and $\theta$ for which the final state is entangled for Compton scattering with Werner input. (b) Logarithmic negativity as a function of $p$ and $\theta$ for Compton scattering with Werner input.}%
\end{figure}

Let us now show that even partial 
filtering of the input state allows one to recover substantial entanglement from this process. To this aim, let us consider the input Werner state obtained by projecting an unpolarised state on the symmetric subspace, given by $\rho_{-\infty} = 
\frac13 (\ket{LL}\bra{LL}+\ket{\psi^+}\bra{\psi^+}+\ket{RR}\bra{RR})$ \cite{Werner1989}. Switching to  this input one gets regions of output entanglement as is shown in Fig.~\ref{fig:egegWernerEMAP}.

As shown in Fig.~\ref{fig:egegWernerLNEG},
Compton scattering is actually capable to generate 
substantial degrees of entanglement for partially filtered inputs. 
Interestingly, in the limit where the scattering angle approaches the divergent back-scattering value $\theta=\pi$ and the COM momentum diverges, the Bell state $\ket{\psi^+}$ may be approached arbitrarily well.

%%%%%%%%%%%%%%%%%%%%%%%%%%%%%%%%%%%%
\subsubsection{Pure input}

Finally, let us also look at what happens if we use a pure beam of states with helicity $\ket{LL}$ as the initial state.
In this case, the final system is {\em always} entangled.

\begin{figure}[b!]
    \centering
    \includegraphics[scale=0.4]{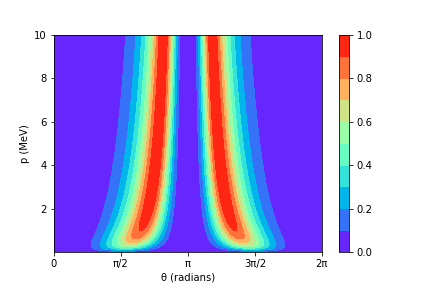}
    \caption{Plot of the logarithmic negativity as a function of $p$ and $\theta$ for Compton scattering (Pure $\ket{LL}$ beam).}
    \label{fig:egegLLLNEG}
\end{figure}

The logarithmic negativity of the system is then shown in Fig.~\ref{fig:egegLLLNEG}. 
The system is maximally entangled in the limit of $\theta\rightarrow \pi$ and $p\rightarrow\infty$, where the output state approaches arbitrarily well the Bell state $\ket{\phi^-}$. 
Even at finite energies and away from the infrared divergence, though, maximal entanglement is approached 
remarkably well. For instance, for  $\theta=3\pi/4$ and $p=3.7 \,{\rm MeV}$ a logarithmic negativity $E_{\mathcal N} \approx 0.98$ is achieved; in this case, the $u$-channel dominates the scattering statistics, and is responsible for the outgoing entanglement.

Thus, we have shown that maximal entanglement can be approached arbitrarily well in Compton scattering too.

\section{Conclusions and outlook}

The systematic approach for analysing 
entanglement in two-particle collisions 
presented in this paper allowed us to cover tree-level entanglement in QED on general grounds, encompassing arbitrary state preparations and dynamical conditions. This brought to light very interesting, previously unknown cases, such as the possibility to reach maximal entanglement in Bhabha scattering and, through the filtering of the initial helicity, Compton scattering.
In such cases, we were able to analyse the generation of entanglement in detail, and to identify optimal dynamical conditions to that purpose.
Our results apply directly to other leptons in the QED sector and our method
can be extended to other QFTs and to higher perturbation levels. 

Besides allowing us to determine the exact dynamical regions of tree-level entanglement  (in analytical form for Møller scattering), 
our study brought to light further mechanisms for the generation of entanglement in QED besides those already highlighted for the generation of exact maximal entanglement in \cite{Cervera-Lierta2017}. In particular, besides the already known single $s$-channel and $t$- and $u$- channel interference generations, we have shown that maximal or `virtually maximal' (i.e., maximal to all practical purposes, given the noise at play) entanglement can originate in all situations where one dominating channel (which could be a $t$- or $u$- one too) leads to balanced superpositions of helicity states. As heuristic rule of thumb, one would be tempted to argue that processes with equal output particle masses are generally more favourable for the onset of such maximal entanglement conditions (and thus no maximally entanglement was found for electron-muon scattering), although this is contradicted by the regions of high entanglement observed for Compton scattering with filtered initial states.

One potential perspective of systematic studies like the present one arises in the context of the current wave of proposed experiments aimed at establishing the quantum nature of gravity, such as in the BMV experiment \cite{Bose2017,Marletto2017,Christodoulou2019,Belenchia2016,Chevalier2020,Kamp2020,Nguyen2020,marshman20,Carney2021,Howl2021,Liu2021,Datta2021,Kent2021,bose22,biswas22,Galley2022,Anastopoulos2022,Danielson2022}. Such proposals suggest to leverage the occurrence of quantum entanglement -- ideally as witnessed by the violation of Bell-like inequalities -- to certify the quantum nature of a fundamental force, i.e., its coherent action at the Hilbert space level. Of course, as such, these tests are primarily aimed at gravity (although they may be relevant to the investigation of other effects, such as those related to Casimir interactions or to the occurrence of axions \cite{barker22}). However, the detection of entanglement in any fundamental quantum field theory, the optimisation of the dynamical conditions for its emergence and a detailed understanding of the mechanisms behind its generation are certainly interesting fundamental questions, whose answers hold potential for wider implications.

%%%%%%%%%%%%%%%%%%%%%%%%%%%%%
\section{Acknowledgements}

We thank the organisers of the 24th Symposium on Quantum Matter and High-Energy Physics, Leeds, 6-8 June 2022, for their invitation and the ensuing discussion and suggestions. 

\bibliographystyle{apsrev}
\bibliography{library}

\end{document}